\documentclass[a4paper,twoside]{article}
\newcommand{\affil}[1]{$^{\mathrm #1}$}
\usepackage[authoryear]{natbib}
\bibpunct{(}{)}{;}{a}{}{,}
\usepackage{times}
\usepackage{graphicx}
\graphicspath{{./plots/}}
\usepackage[fleqn]{amsmath}
\newcommand{\romn}[1] {{\mathrm #1}}
\newcommand\fs{\hbox{$.\!\!^{\romn s}$}}
\newcommand\farcs{\hbox{$.\!\!^{\prime\prime}$}}
\newcommand\arcmin{\hbox{$^\prime$}}
\newcommand\degr{\hbox{$^\circ$}}
\date{} 
\title{\large\bf\flushleft Application of a Bayesian Method to
  Absorption Spectral-Line Finding in Simulated ASKAP Data}
\author{\parbox{\textwidth}{\flushleft
\vspace{-0.5cm}
{\it J.~R. Allison\affil{A,D}, E.~M. Sadler\affil{A,B}, and M.~T. Whiting\affil{C}}\\
\vspace{0.4cm}
{\small \affil{A}\,Sydney Institute for Astronomy, School of Physics A28, University of Sydney, NSW 2006, Australia}\\
{\small \affil{B}\,ARC Centre of Excellence for All-sky Astrophysics (CAASTRO)}\\
{\small \affil{C}\,CSIRO Astronomy \& Space Science, P.O. Box 76, Epping NSW 1710, Australia}\\
{\small \affil{D}\,Corresponding author. Email: jra@physics.usyd.edu.au}}}
\begin{document}
\twocolumn[
\begin{changemargin}{.8cm}{.5cm}
\begin{minipage}{.9\textwidth}
\vspace{-1cm}
\maketitle
\small{\bf Abstract:} The large spectral bandwidth and wide
field of view of the Australian SKA Pathfinder radio telescope will
open up a completely new parameter space for large extragalactic HI
surveys. Here we focus on identifying and parametrising HI
absorption lines which occur in the line of sight towards strong radio
continuum sources.  We have developed a method for simultaneously
finding and fitting HI absorption lines in radio data by using
multi-nested sampling, a Bayesian Monte Carlo algorithm. The method
is tested on a simulated ASKAP data cube, and is shown to be reliable
at detecting absorption lines in low signal-to-noise data without the
need to smooth or alter the data. Estimation of the local Bayesian
evidence statistic provides a quantitative criterion for assigning
significance to a detection and selecting between competing analytical
line-profile models.


\medskip
\medskip
\end{minipage}
\end{changemargin}
]
\small

\section{Introduction}

The Australian Square Kilometre Array Pathfinder's
\citep[ASKAP;][]{Deboer:2009} large spectral bandwidth and wide field
of view will dramatically improve our ability to conduct large-area
galaxy surveys in the 21\,cm line of neutral hydrogen
\citep{Johnston:2007}.

The ASKAP HI All-Sky Survey (WALLABY Science Survey Proposal;
Koribalski \& Staveley-Smith
2009\footnote{http://www.atnf.csiro.au/research/WALLABY})
will cover 75 of the sky
($-90\,\mathrm{deg}<\mathrm{Dec.}<+30\,\mathrm{deg}$) at a spatial
resolution of approximately 30\,arcsec and velocity resolution of
approximately 4\,km\,s$^{-1}$. With an integration time of 8\,h per
pointing (assuming a system temperature of 50\,K) the survey will
allow us to examine the HI properties and large-scale distribution of
$\sim$500,000 galaxies out to a redshift of 0.26 (equivalent to a
look-back time of approximately 3\,Gyr).

ASKAP will also be a powerful instrument for carrying out blind HI
absorption-line surveys using background radio continuum sources. The
advantage of absorption-line surveys is that their sensitivity depends
only on the brightness of the background source, making it possible to
probe the neutral gas content of individual galaxies at redshifts
where the HI emission line is too weak to be detectable.

The ASKAP First Large Absorption Survey in HI (FLASH Science Survey
Proposal; Sadler et
al. 2009\footnote{http://www.physics.usyd.edu.au/sifa/Main/FLASH})
will search for HI and OH absorption features in two redshift ranges
($0<z<0.26$ and $0.5<z<1.0$) using bright background continuum sources
from the existing SUMSS \citep[][]{Mauch:2003} and NVSS
\citep[][]{Condon:1998} catalogues, both of which have an angular
resolution of 45\,arcsec. This amounts to a targeted search of over
150,000 sightlines to background continuum sources, an increase of
more than two orders of magnitude over the total number of sightlines
probed in all previous HI absorption-line surveys with radio
telescopes. In the lower ($0<z<0.26$) redshift range, the same ASKAP
data are used for the FLASH and WALLABY surveys, making it possible to
cross-compare emission- and absorption-line measurements of local
galaxies.

FLASH will search all ASKAP HI data cubes for absorption lines at the
positions of radio continuum sources with flux densities above 50\,mJy
in the 1.4\,GHz NVSS and 843\,MHz SUMSS surveys. Since the positions
of these background continuum sources are already known, the
`source-finding problem' for FLASH is reduced to the need for a
reliable `line-finding' algorithm which can be efficiently applied at
a large number of pre-determined positions on the sky. When
characterising the detected lines, we want to obtain a reliable
analytical model of the line profile and distinguish between competing
models, even in the low signal-to-noise (SNR) regime.

A robust quantitative method of selecting between competing models,
and measuring the significance of a detection, is provided through the
calculation of the Bayesian evidence statistic. This method is already
being used for a range of other low SNR astrophysical scenarios,
including model fitting to observations of the Sunyaev-Zel'dovich
Effect \citep[see e.g.][]{Marshall:2003, Feroz:2009a, Allison:2011a},
where we are interested in comparing between competing models for a
redshift-independent observable.

In this paper we present the application of an existing Bayesian Monte
Carlo algorithm to the problem of assigning significance to the
detection and modeling of HI absorption lines in a simulated ASKAP
data cube. Unless otherwise stated, all errors refer to the 68.3\,\%
interval.

\section{Simulated data}

The expected properties of a full ASKAP data cube include a
30\,deg$^{2}$ field of view and 300\,MHz of bandwidth with 16,384
channels, corresponding to an HI velocity resolution of
12\,km\,s$^{-1}$ at 800\,MHz. Present computing limitations meant that
it was only possible to simulate 1024 spectral channels over the full
30\,deg$^{2}$ ASKAP field, equating to 18\,MHz of spectral bandwidth.
A simulated ASKAP--FLASH data cube covering the redshift range
$0.76<z<0.792$ was released by the ASKAP computing group in May 2011,
and details were made publically available
online\footnote{http://www.atnf.csiro.au/people/Matthew.Whiting/ASKAPsimulations.php}.

\begin{figure}
\centering
{\includegraphics[width = 1.00\columnwidth]{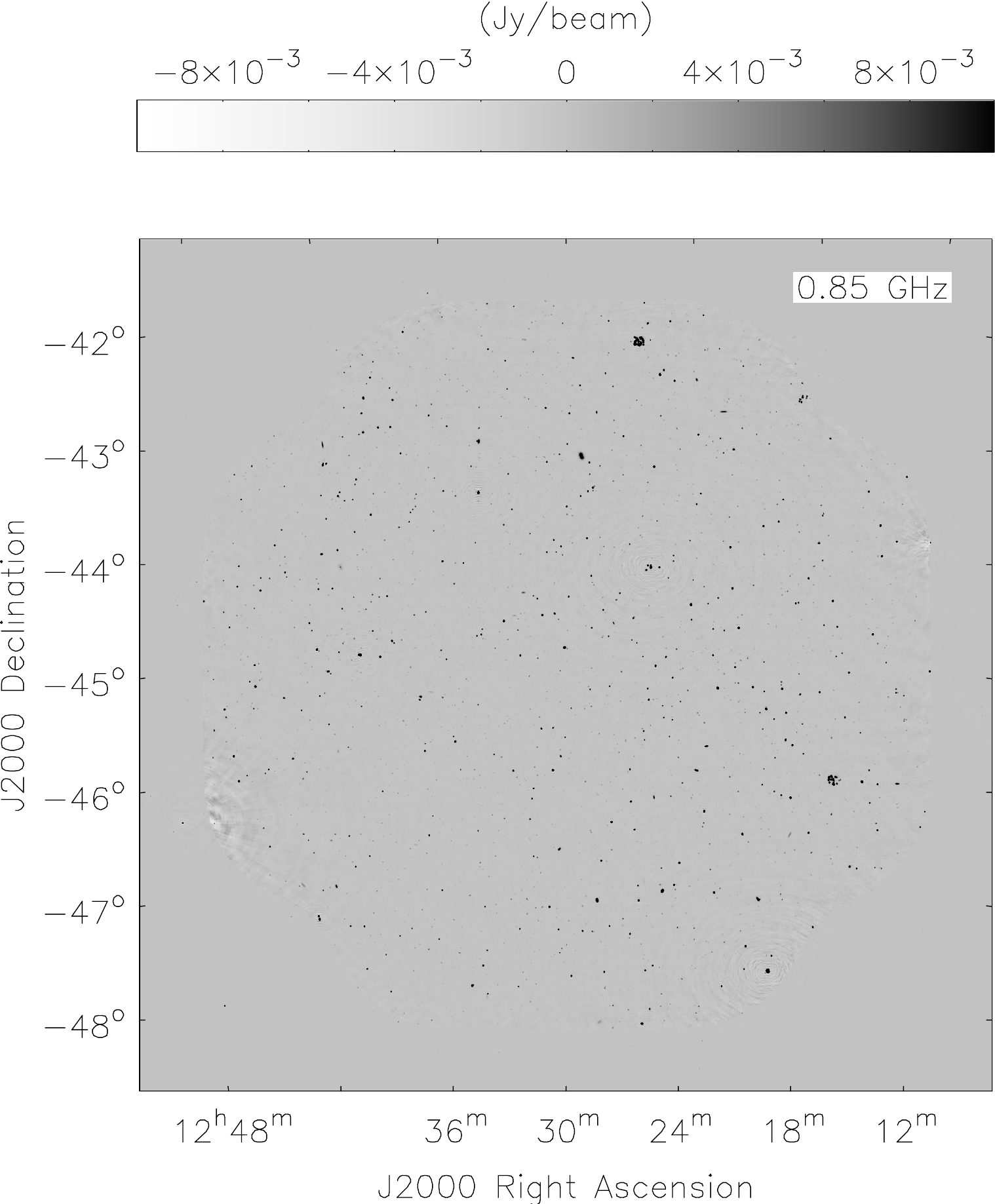}}
{\includegraphics[width = 1.10\columnwidth]{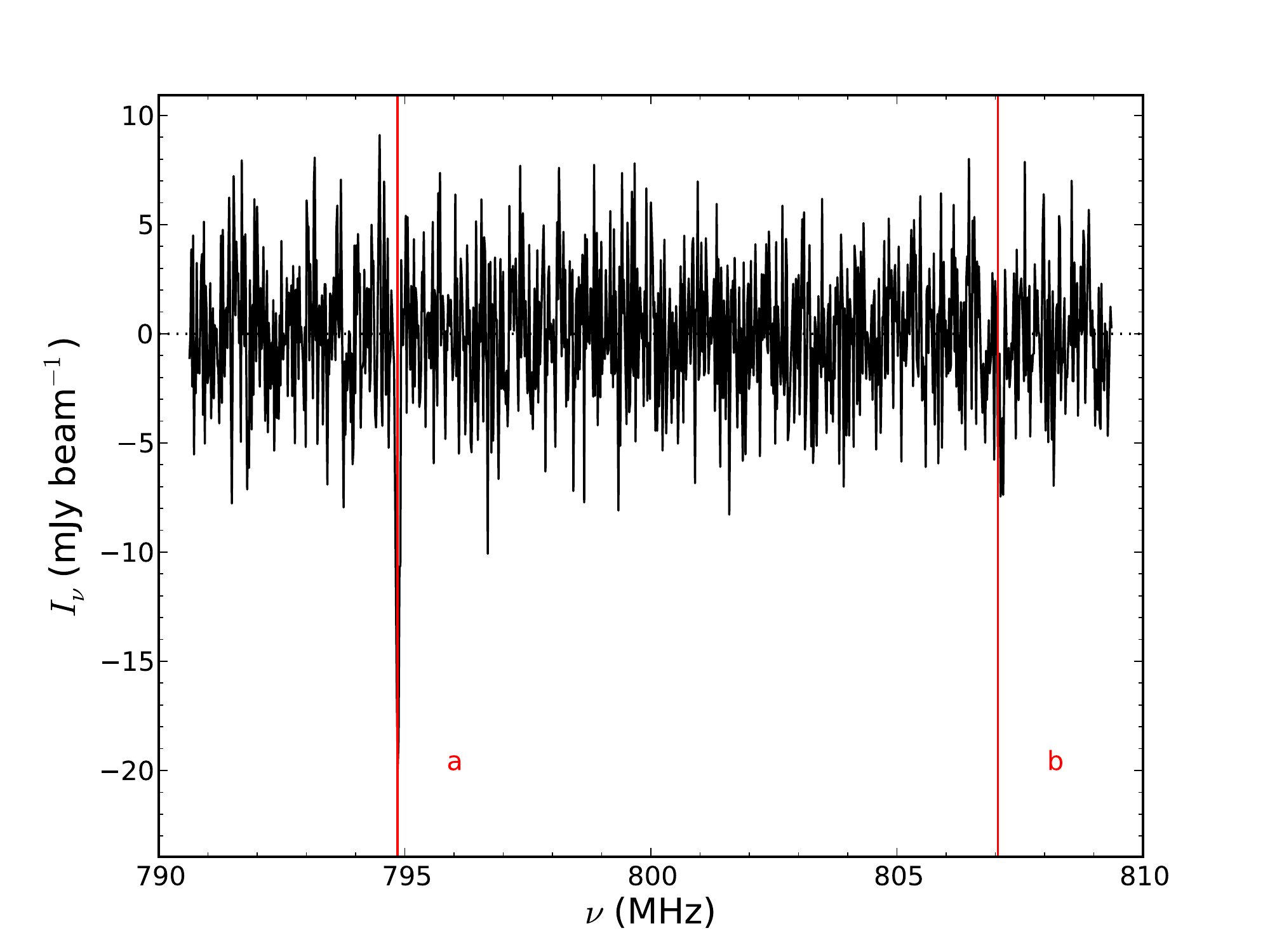}}
\caption{\emph{Top}: A simulated 850MHz continuum ASKAP image based on
  the semi-empirical SKADS simulation by \cite{Wilman:2008}. The
  maximum pixel value has been limited to 0.01\,Jy\,beam$^{-1}$ to
  prevent the brightest sources from dominating the
  image. \emph{Bottom}: An example continuum-subtracted spectrum
  extracted from the ASKAP--FLASH simulated data cube at the position,
  $\mathrm{RA(J2000)} = 12^{\mathrm{h}} 28^{\mathrm{m}} 26\fs86$ and
  $\mathrm{Dec.(J2000)} = -47{\degr}03{\arcmin}31{\farcs}50$, of a
  source of flux density $S_{800}$ = 198.7\,mJy. The \emph{red lines}
  indicate the positions of two HI absorption-line components present
  in the spectrum. Component \emph{a} is clearly visible above the
  noise level.}
\label{figure:simulation}
\end{figure}

The FLASH simulation included both spectral-line and continuum
information, and the basic steps were:

\begin{enumerate}
\item Create a realistic continuum sky simulation at 850 MHz, using
  the semi-empirical SKADS simulation by \cite{Wilman:2008} and an
  assumed integration time of two hours per pointing (see
  Figure\,\ref{figure:simulation}).

\item `Paint in' a grid of Gaussian HI absorption-line profiles
  covering a range in velocity full width at half maximum (FWHM) and
  peak optical depth $\tau$.  S/N calculations indicate that only
  sources stronger than about 50\,mJy\,beam$^{-1}$ are realistic
  targets for the FLASH survey (with a planned observing time of two
  hours per ASKAP pointing), but sources with flux densities down to
  10\,mJy\,beam$^{-1}$ had HI absorption lines added in the simulation
  so that our line-finding method could be tested in the low S/N
  regime.
\end{enumerate}

To provide a useful test of line-finding algorithms, the number of
absorption lines inserted into the simulated data is far higher than
the number that we would expect to see in a real ASKAP data cube.  In
total 600 lines (each with a single Gaussian profile) were inserted
into the simulated cube.  They spanned a redshift range
$0.76<z<0.792$, with optical depths $0.01<\tau<0.30$ and velocity
widths between 5 and 80\,km\,s$^{-1}$. Not all of these lines are
expected to be detectable in the final simulated data cube.

The continuum and spectral-line datasets were kept separate to mimic
the effects of continuum subtraction, since the capability to do this
in the ASKAP pipeline had not yet been fully implemented.

\section{Method}

To search for HI absorption from neutral gas in distant galaxies we
target sightlines towards known bright continuum sources, since the
detection probability for an HI absorption line is independent of
redshift but increases with the brightness of the background continuum
source.  The blind aspect of the survey arises from searching for
absorption dips in the spectral domain, so we do not need a
3-dimensional source finder, such as
\textsc{Duchamp}\footnote{http://www.atnf.csiro.au/people/Matthew.Whiting/Duchamp/}
\citep[][Whiting et al. in prep.]{Whiting:2008}. Instead, we need a
tool that detects any spectral lines, quantifies their properties
based on an analytical model, and provides an estimate of the
detection significance. Standard $\chi^{2}$ minimisation and residual
inspection have previously been used to fit parametrised Gaussian
models in HI absorption surveys \citep[see e.g.][]{Gupta:2010,
  Kanekar:2009}, and the analysis outlined in this work uses a
generalised extension of those methods. In the following sections we
discuss a Bayesian approach to the one-dimensional line-finding
problem.

\subsection{Spectra extraction}
One-dimensional spectra are extracted from the simulated ASKAP data
cube using a scripted \textsc{Python} routine in the
\textsc{Casa}\footnote{http://casa.nrao.edu/} data reduction package.
An input catalogue of the 435 continuum sources that contain simulated
HI absorption lines is used to provide the known positions from which
the spectra are extracted. Extraction is performed at the centre
position of each source using the task \textsc{Imval} in
\textsc{Casa}. Each source is indexed based on its flux density at
$800\,$MHz, ordered in descending value.  Figure
\ref{figure:simulation} shows an example of an extracted spectrum from
a continuum source, in which there are two HI absorption-line
components. One of the components is clearly visible by eye at
794.9\,MHz ($z = 0.787$), while the other is buried within the noise
at 807.0\,MHz ($z = 0.760$). The spectral data are stored as
individual data files for each continuum source with information on
the frequency, brightness and uncertainty. The uncertainties in the
data are estimated based on the median of the absolute deviations from
the median value (MADFM). This statistic is a more robust estimator of
the true uncertainty than the RMS when a strong signal is present in
the data. For Gaussian distributed data, the true standard deviation
is estimated by multiplying the MADFM statistic by a factor of
1.4826042. The spectra are also stored in Flexible Image Transport
System \citep[FITS,][]{Wells:1981} format, so that they are compatible
for use with the \textsc{Duchamp} source finder.

\begin{figure*}
\centering
\includegraphics[width = 2.00\columnwidth]{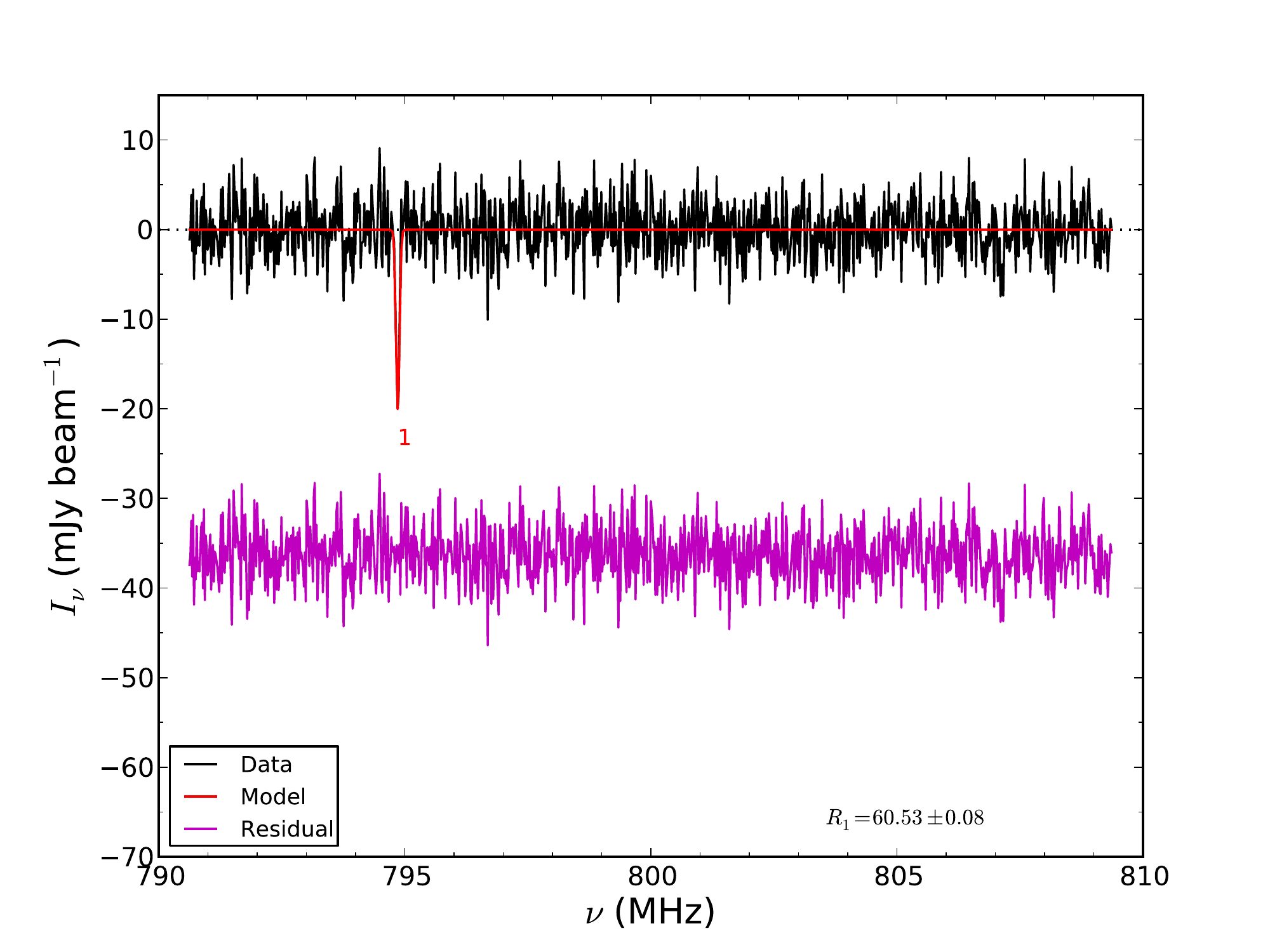}
\includegraphics[width = 2.25\columnwidth]{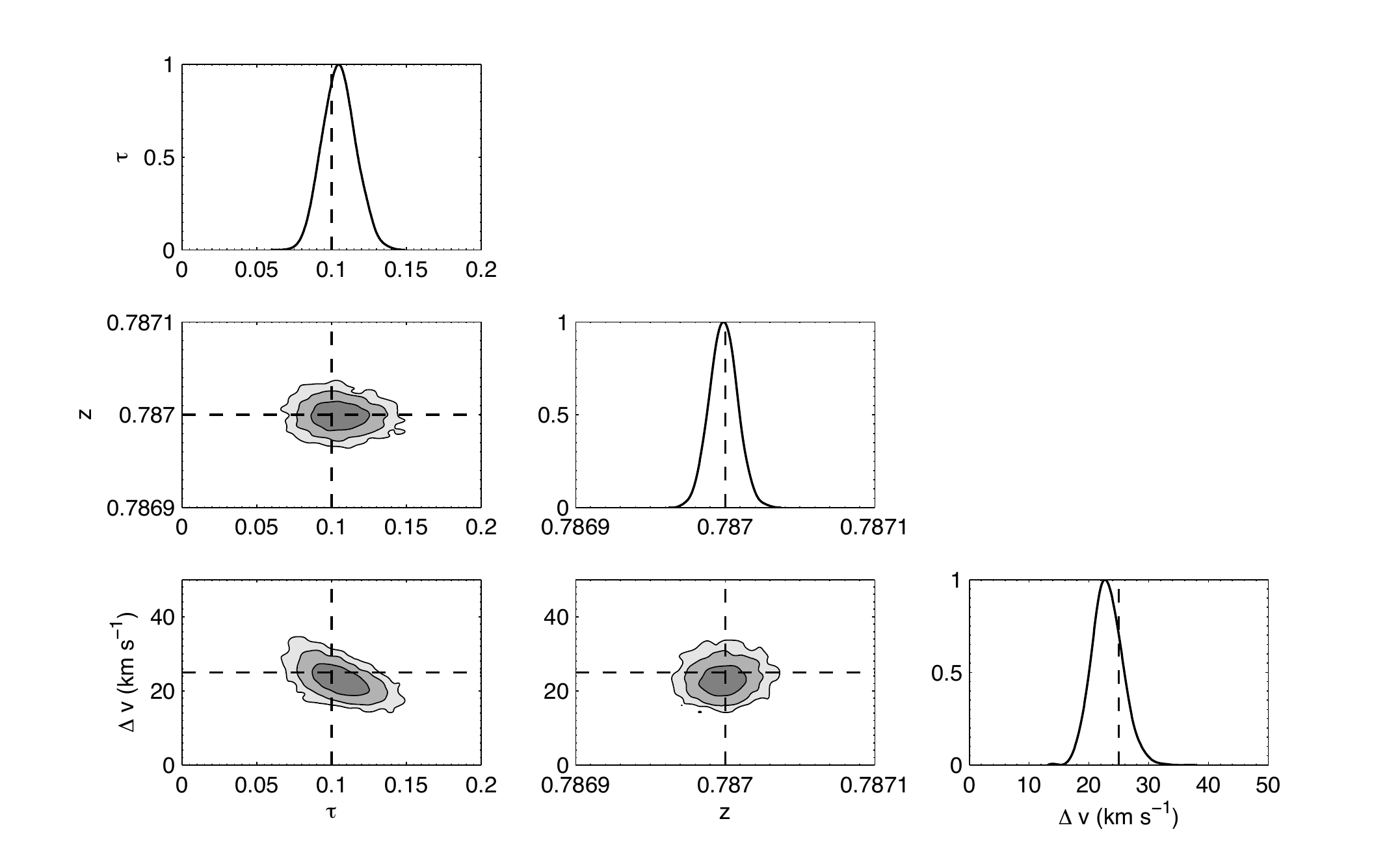}
\caption{\emph{Top}: Example of spectral-line fitting to a simulated
  ASKAP source ($\mathrm{RA} = 12^{\mathrm{h}} 28^{\mathrm{m}}
  26\fs86$, $\mathrm{Dec.} = -47{\degr}03{\arcmin}31{\farcs}50$,
  $S_{800}$ = 198.7\,mJy). One of the absorption-line components is
  detected , while the other is hidden in the simulated noise. The
  residual has been plotted with an offset from the frequency axis for
  clarity. \emph{Bottom}: Estimate of the marginialised posterior
  probabilities for absorption-line parameters from the example
  detected source. The parameters displayed are the peak optical depth
  ($\tau$), redshift ($z$) and velocity FWHM ($\Delta v$). The
  grey-scale represents the 68.3, 95.4, and 99.7\% intervals. The
  dashed lines represent the input catalogue values.}
\label{figure:example_detection}
\end{figure*}

\subsection{Bayesian inference}

We fit analytical models to the extracted spectral data using Bayesian
inference.  The posterior (or joint) probability for a set of model
parameters ($\boldsymbol{\theta}$), given the data ($\boldsymbol{d}$)
and the model hypothesis ($\mathcal{M}$), can be calculated from
Bayes' theroem,
\begin{equation}\label{equation:bayes_theorem}
  \mathrm{Pr}(\boldsymbol{\theta}|\boldsymbol{d},\mathcal{M}) =
  {\mathrm{Pr}(\boldsymbol{d}|\boldsymbol{\theta},\mathcal{M})\mathrm{Pr}(\boldsymbol{\theta}|\mathcal{M})\over\mathrm{Pr}(\boldsymbol{d}|\mathcal{M})}.
\end{equation}
The probability of the data given the model parameters, known as the
likelihood, can be calculated based on assumptions about the
distribution of the uncertainty in the data. If the data set is large
and therefore quasi-continuous (such as the thermal noise generated in
radio instrumentation), one can approximate the likelihood by the form
given for Gaussian multivariate data \citep[see e.g.][]{Sivia:2006}
\begin{flalign}\label{equation:Gaussian_likelihood}
  L & \equiv \mathrm{Pr}(\boldsymbol{d}|\boldsymbol{\theta},\mathcal{M}) & \nonumber \\ 
  & = {1\over\sqrt{(2\pi)^{N}|\mathbf{C}|}}\exp{\left[-{(\boldsymbol{d}-\boldsymbol{m})^\mathrm{t}\mathbf{C}^{-1}(\boldsymbol{d} - \boldsymbol{m})\over2}\right]}, &
\end{flalign}
where $N$ is equal to the size of $\boldsymbol{d}$, $\mathbf{C}$ is
the covariance matrix of the data, $|\mathbf{C}|$ is the determinant
of the covariance matrix and $\boldsymbol{m}$ is the vector of model
data. In the special case where the variance in the data is a constant
($\sigma^{2}$) and uncorrelated, the above expression for the
likelihood reduces to
\begin{equation}
  L = {1\over\sigma^{N}\sqrt{(2\pi)^{N}}}\exp{\left[-{\sum_\mathrm{i}(d_\mathrm{i} - m_\mathrm{i})^{2}\over2\sigma^{2}}\right]}.
\end{equation}
The probability of the parameter values given the model hypothesis,
$\mathrm{Pr}(\boldsymbol{\theta}|\mathcal{M})$, is often known as the
prior probability and encodes information about the parameter values
\emph{a priori}. For example, consider the situation where the
frequency position of an intervening HI absorber has been relatively
well constrained from previous observations. If we trust these
observations we might then choose to apply a normal prior to the
spectral-line position based on the known level of uncertainty. We
would otherwise apply uninformative priors to the parameters if we
were are previously unaware of their value. Uninformative priors are
typically uniform in either linear space (for location parameters) or
logarithmic space (for scale parameters, known as Jeffery's prior).

The normalisation of the posterior probability in
Equation\,\ref{equation:bayes_theorem} is equal to the probability of
the data given the model hypothesis and is referred to throughout this
work as the evidence. The evidence is calculated by marginalising the
product of the likelihood and prior distributions over the model
parameters. This is given by
\begin{flalign}\label{equation:evidence}
  E &\equiv \mathrm{Pr}(\boldsymbol{d}|\mathcal{M}) & \nonumber\\
  & = \int{\mathrm{Pr}(\boldsymbol{d}|\boldsymbol{\theta},\mathcal{M})\mathrm{Pr}(\boldsymbol{\theta}|\mathcal{M})\mathrm{d}\boldsymbol{\theta}}, &
\end{flalign}
which follows from the relation given by
Equation\,\ref{equation:bayes_theorem} and that the integrated
posterior is normalised to unity. When the model hypothesis provides a
good fit to the data the likelihood peak will have a large value, and
hence the model hypothesis will have a large associated evidence
value. However if the model is over-complex then there will be large
regions of low likelihood within the prior volume, thus reducing the
evidence value for this model, in agreement with Occam's
razor. Estimation of the evidence is often key in providing a tool for
selecting between competing models.

\subsection{Application to spectral-line finding}

In the approach presented in this work we wish to ask the question: Do
the data warrant a model hypothesis that includes the presence of a
spectral-line of a given form, in preference to a model with no
spectral-lines at all? In the case of the simulated ASKAP--FLASH data
the underlying signal is know to be a single Gaussian component, with
all the continuum signal perfectly subtracted. Hence we use a
spectral-line model hypothesis that is given by a single Gaussian of
the form
\begin{equation}
  I_\mathrm{\nu}  = I_\mathrm{\nu,peak}\exp{\left[-4\ln{(2)}{(\nu - \nu_\mathrm{peak})^{2}\over (\Delta \nu)^{2}}\right]},
\end{equation}
where the set of model parameters $\boldsymbol{\theta}$ consist of the
characteristic peak value $I_\mathrm{\nu,peak}$, the spectral position
$\nu_\mathrm{peak}$, and the FWHM of the spectral line $\Delta
\nu$. We test this single Gaussian spectral-line model against the
null hypothesis of a model containing no spectral line at all. In the
case of perfectly continuum subtracted data we expect there to be no
signal (i.e. $m_\mathrm{i, null} = 0$ for all i) and so the likelihood
of the data for the null model reduces to
\begin{equation}
  L_\mathrm{null} = {1\over\sigma^{N}\sqrt{(2\pi)^{N}}}\exp{\left[-{\sum_\mathrm{i}(d_\mathrm{i})^{2}\over2\sigma^{2}}\right]}.
\end{equation}

We are simulating a blind absorption survey and so use uninformative
priors for all of the parameters in our spectral-line model (see
Table\,\ref{table:priors}). The line-depth prior is set by the
physical limit of the brightness of each source. We can also search
for emission by reversing the sign of the line-depth prior and instead
consider positive values. The spectral position is limited to the
range of frequencies recorded by the data. The prior range for the
FWHM of the spectral line correspond to a velocity range of $\sim$
0.65 -- 650\,km\,s$^{-1}$ at 800\,MHz, which are considered to be
physically reasonable limits.

We use the ratio of the probabilities for model hypotheses given the
data,
\begin{equation}
  {\mathrm{Pr}(\mathcal{M}_{1}|\boldsymbol{d})\over  \mathrm{Pr}(\mathcal{M}_{2}|\boldsymbol{d})} = {\mathrm{Pr}(\boldsymbol{d}|\mathcal{M}_{1})\over  \mathrm{Pr}(\boldsymbol{d}|\mathcal{M}_{2})}  {\mathrm{Pr}(\mathcal{M}_{1})\over  \mathrm{Pr}(\mathcal{M}_{2})} = {E_{1}\over  E_{2}}  {\mathrm{Pr}(\mathcal{M}_{1})\over  \mathrm{Pr}(\mathcal{M}_{2})}, \\
\end{equation}
to quantify the relative significance of the Guassian spectral-line
versus no-line model. The ratio
${\mathrm{Pr}(\mathcal{M}_{1})/\mathrm{Pr}(\mathcal{M}_{2})}$ encodes
our prior belief that one hypothesis is favoured over another. Since
we assume no prior information on the presence of spectral lines, this
ratio is equal to unity and so the above selection criterion is then
given by the ratio of the evidences. We define the quantity
\begin{equation}
  R \equiv \ln\left({E_\mathrm{Gauss}\over E_\mathrm{null}}\right),
\end{equation}
with values greater than zero indicating the the level of significance
for the Guassian spectral-line detection. Values of $R$ less than zero
indicate that the data do not warrant the inclusion of the Guassian
spectral-line model over the null hypthothesis and so the detection is
rejected.

It should be noted that we haven only chosen to use single Gaussians
to parametrize the absorption lines, which for the case of the
simulated ASKAP--FLASH data is equal to the underlying model. However
the technique can be used for any model parametrization of the
spectral-line profile. The validity of using more complex models for a
given data set can be inferred by comparing the successive evidence
values. Indeed we can follow up a detection using the single Guassian
profile by incrementally increase the number of components and compare
the evidence for each model hypothesis until a best fit is
obtained. The evidence statistic will penalise overly complex models
and so will likely reach an optimised value after a fixed number of
components. The quality of the best-fit model can also be inferred
qualitatively by inspection of the residual spectrum. In addition to
more complex spectral-line models we may also wish to simultaneously
fit to the continuum spectrum, rather than subtracting a best-fit
continuum model prior to analysis. In this case we can compare a
continuum and spectral line to a continuum-only model and therefore
again infer the presence of spectral lines in our data. This has the
benefit of correctly propagating the uncertainties in the continuum
model parameters through to the derived marginalised probability
distributions of our spectral-line model parameters.

\subsection{Implementation}

\begin{table} 
  \centering
\caption{Model parameter priors} 
\begin{tabular}{lcc} 
  \hline
  Parameter & Prior type & Prior range\\ 
  \hline
  $I_\mathrm{\nu,peak}$ & log-uniform & $\pm (0.001$\,mJy -- $I_{800}$) \\
  $\nu_\mathrm{peak}$  & linear-uniform & 790 -- 810\,MHz \\
  $\Delta \nu$ & log-uniform & 0.001 -- 1\,MHz \\ 
  \hline
\end{tabular} 
\label{table:priors} 
\end{table}

Bayesian model fitting is implemented using the existing
\textsc{MultiNest}\footnote{http://ccpforge.cse.rl.ac.uk/gf/project/multinest/}
package developed by \cite{Feroz:2008} and \cite{Feroz:2009b}. This
software uses nested sampling \citep{Skilling:2004} to explore
parameter space and robustly calculate both the posterior probability
distribution and the evidence for a given likelihood function and
prior (provided by the user).

We run \textsc{MultiNest} with multi-modal switched on, whereby
samples are taken of multiple likelihood peaks within parameter space,
thus allowing for multiple absorption lines. For each peak in
likelihood we calculate a local evidence value for the single Gaussian
model, and then compare it with the evidence of a model with no
line. The significance of the Gaussian-line profile for a given data
set, is inferred by the relative value of the local evidence compared
with $E_\mathrm{null}$.  If the local evidence for the single Gaussian
model is less than or equal to $E_\mathrm{null}$ then this
``detection'' is rejected. Following the successful completion of the
nested sampling algorithm, both the multi-modal local evidence values
and the model parameter posterior probability are recorded. In this
work we use simple Message Passing Interface (MPI) to split the
spectral data across multiple processors, however \textsc{MultiNest}
has intrinsic MPI capability and the use of this for ASKAP--FLASH will
be investigated in future work.

The method described in this work, whereby we infer the probability of
the spectral model given the data, is a forward approach to the
problem and hence we do not apply a smoothing kernel to the spectral
data. To do so would introduce assumptions about the underlying signal
in the data and therefore introduce false detections into the results,
which would be indistinguishable from true detections.

\section{Results and discussion}

\subsection{Output from the line-finder}

Figure\,\ref{figure:example_detection} shows an
example of the output from line detection in a simulated ASKAP
spectrum. In this example one of the two absorption-line components
known to be present in the spectrum has been detected above the noise
and the posterior probability for the Gaussian parameters
estimated. The second absorption-line component at 807.0\,MHz ($z =
0.760$), while having a relatively wide FWHM of $\Delta v =
80$\,km\,s$^{-1}$, has a low optical depth of $\tau = 0.02$ and so was
not detected above the noise. It is clear from the residual spectrum
that no other lines are present above the noise level.

For this example spectrum we calculate that $R = 60.53\pm0.07$,
indicating that the Gaussian-line model is significantly favoured. The
marginalised posterior probability distribution for each parameter is
shown in the lower plot in Figure\,\ref{figure:example_detection} and
are reasonably Gaussian in shape. The 2-dimensional contours indicate
the correlation between parameters. There is no apparent correlation
between the peak optical depth and redshift, or between the FWHM and
redshift.  There is some anti-correlation between the peak optical
depth and FWHM of the line, indicating conservation of the integrated
optical depth.

It has been noted that the simulated absorption catalogue was
constructed based on a single Gaussian-line model. However when
analysing real ASKAP--FLASH data we will have to make an assumption
about the analytical form of the line profile. Calculation of the
Bayesian evidence statistic provides us with a global likelihood for
selecting between competing models. If, for example we choose to
parametrise the data using a Lorentizian-line model (including the
same uniform priors used for the Gaussian-line model) then we obtain a
value of $R = 56.88\pm0.07$.  In this case the evidence again rejects
the no-line model, but favours the Gaussian-line over the
Lorentizan-line model.

\begin{figure}
\centering
\includegraphics[width = \columnwidth]{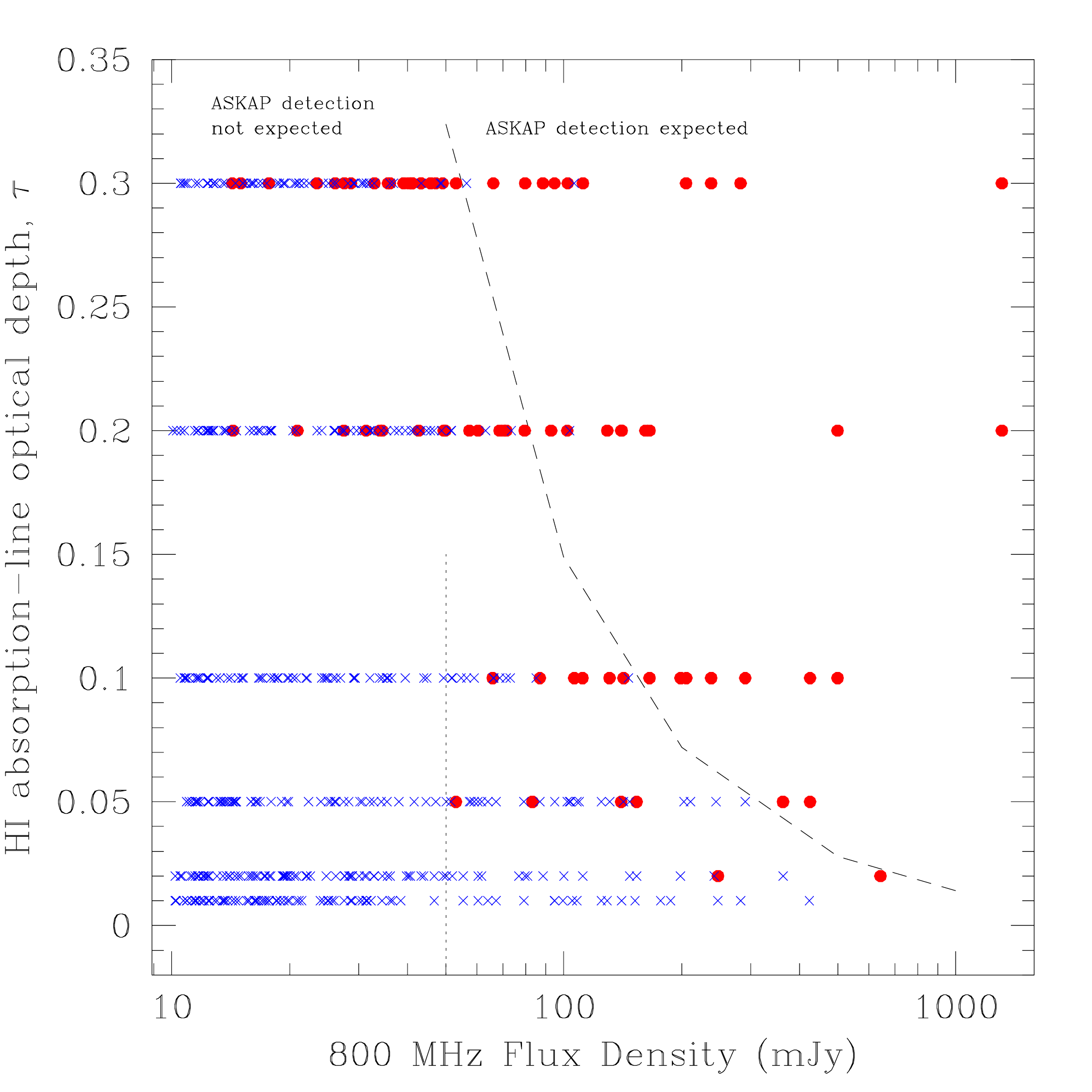}
\includegraphics[width = \columnwidth]{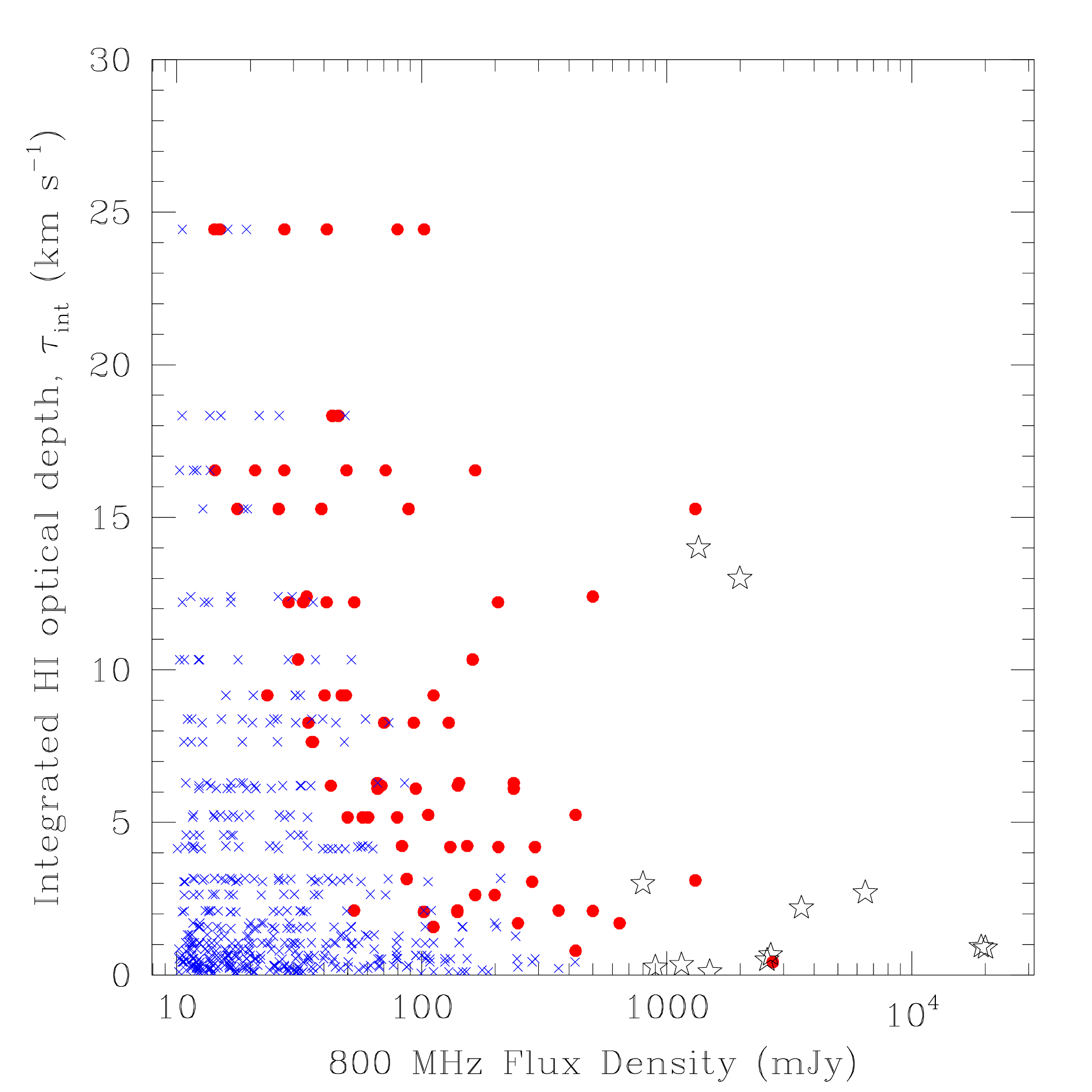}
\caption{The results of running our Bayesian line finder on the
  simulated ASKAP data cube. \emph{Red circles} and \emph{blue
    crosses} represent detected and undetected lines respectively.
  \emph{Top:} Lines of different peak optical depth $\tau$ as a
  function of the flux density of the background continuum source.
  The \emph{dashed line} represents the cut-off for detectability in a
  2\,hour ASKAP observation as originally estimated by the FLASH team
  (see text). \emph{Bottom:} The velocity integrated HI optical depth
  over the line. The black open stars show published observational
  data points for radio detections of intervening HI absorption-lines
  at redshift $z<1$, taken from Table 1 of \citet{Curran:2008}.}
\label{figure:expected_detections}
\end{figure}

\subsection{Comparison with input catalogue}

Of the 600 absorption-line components painted onto the 435 brightest
sources in the continuum simulation, 60 are found to be at locations
off the edge of the main field of the image and are therefore
discounted from the sample. Of the components located within the
image, there are 3 detections with $R$ less than unity. These
detections have comparable significance to the ten false-positive
detections and are therefore counted as non-detections. Note that the
10 false-positive detections (which include 4 in absorption and 6 in
emission) have very low significance and unphysical velocity widths
and hence can be distinguished from the correct detections.

Of the remaining absorption-line components, 76 are detected above the
noise with $R$ greater than unity, yielding a detection rate of 14\,\%
from a realistic 2-h integration on an ASKAP field.

Figure\,\ref{figure:expected_detections} plots the peak and integrated
HI optical depth versus the 800\,MHz flux density of the background
continuum source for both detected and undetected lines from the
ASKAP-FLASH simulation The dashed line in the first plot shows the
detection limit in peak optical depth $\tau$ originally assumed by the
FLASH team, based on the 5-sigma detection of a line peak in a single
18\,kHz spectral channel in a 2-h ASKAP observation.  We detect almost
all of the sources expected to be found in the real ASKAP data, as
well as some additional weaker lines.  The simulation results
therefore imply that the assumed FLASH detection limit is reasonable,
and may even be slightly conservative.

The plot of integrated optical depth $\tau_\mathrm{int}$ in
Figure\,\ref{figure:expected_detections} also shows the existing
observational data points for HI absorption-lines at $z<1$ from Table
1 of \cite{Curran:2008}.  This plot shows that the FLASH survey should
be able to detect similar HI absorption-lines against continuum
sources which are 10-100 times fainter than those typically probed in
targeted HI absorption-line searches with existing radio telescopes.

\begin{figure}
\centering
\includegraphics[width = \columnwidth]{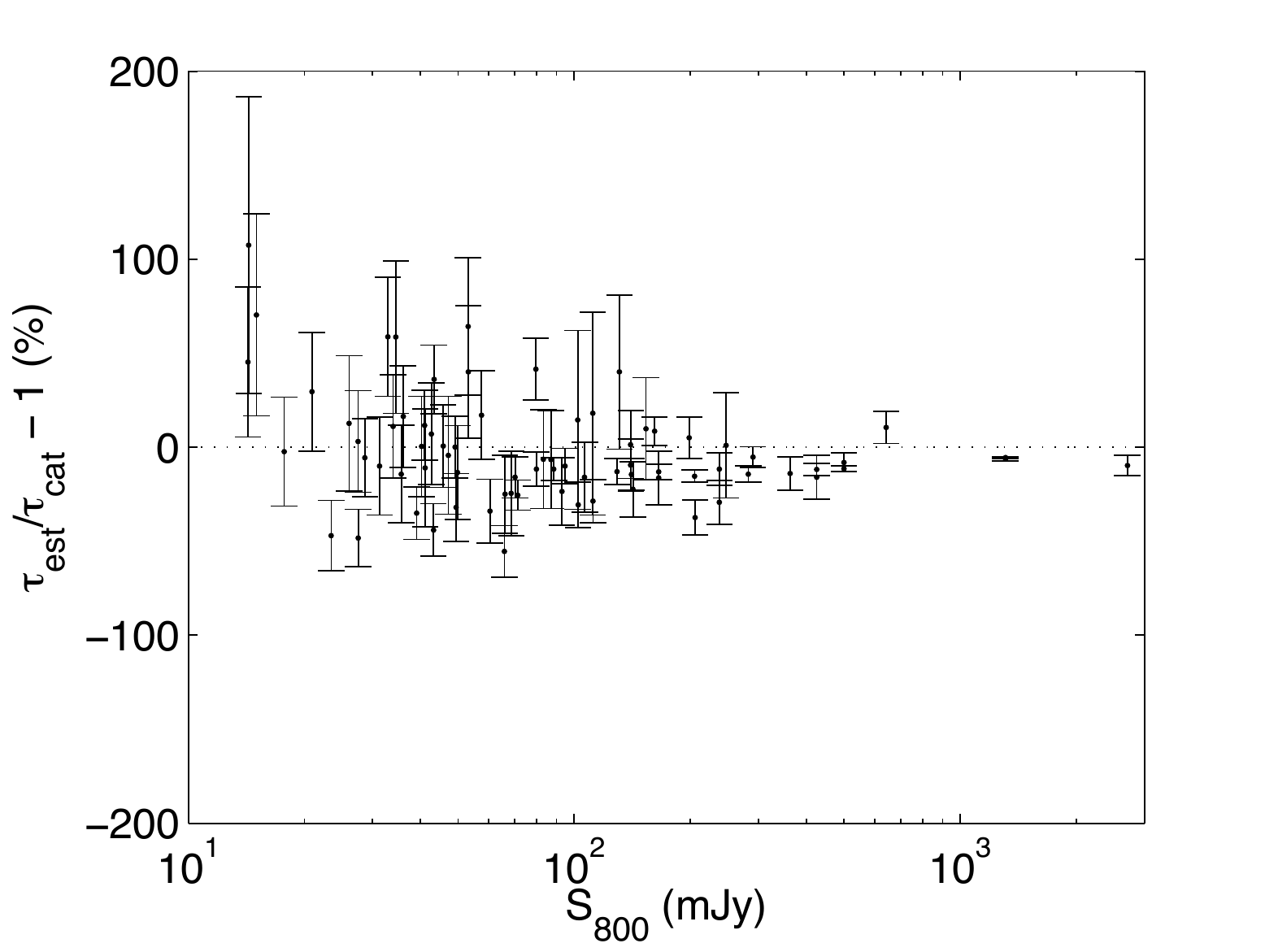}
\includegraphics[width = \columnwidth]{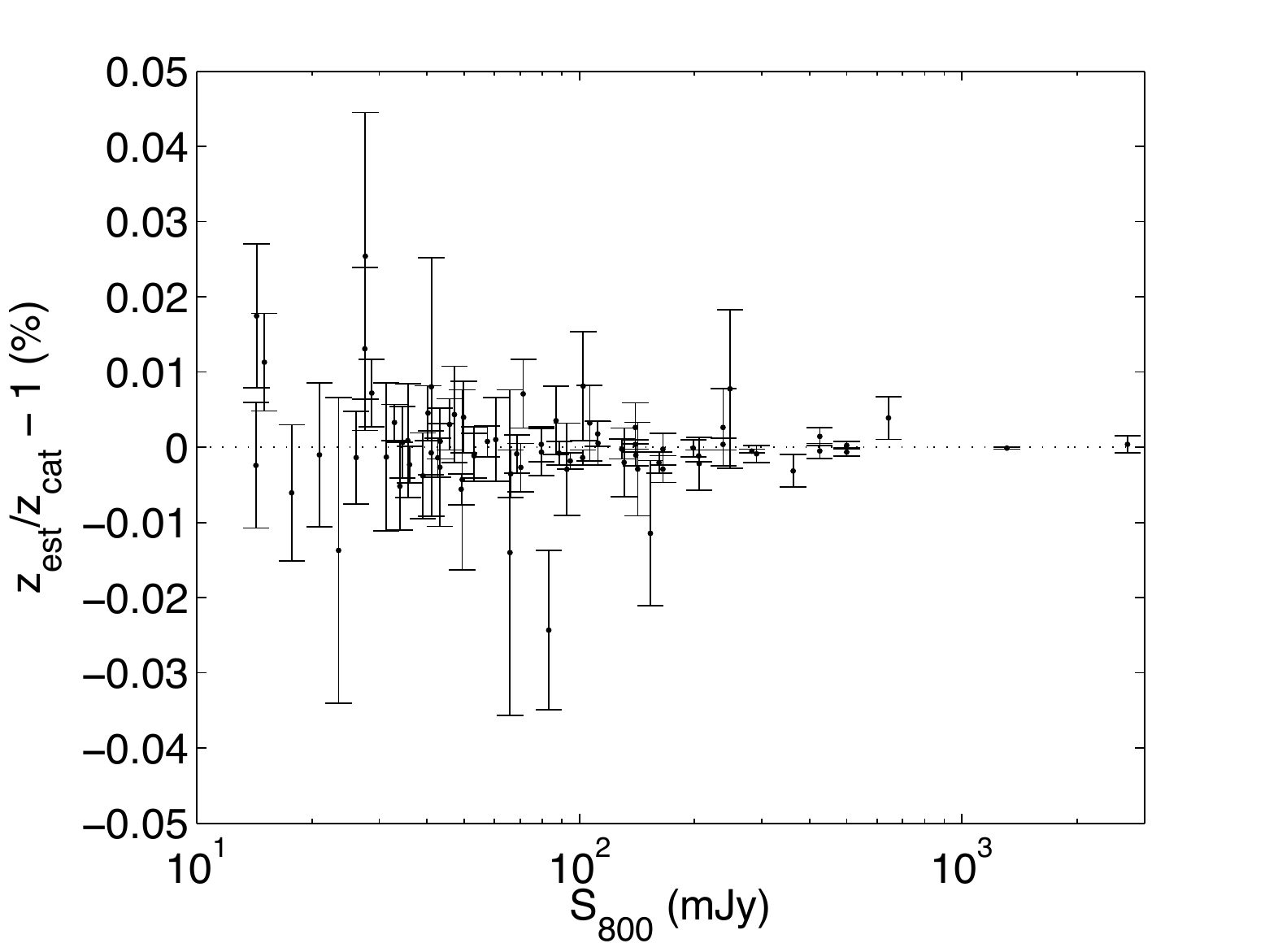}
\includegraphics[width = \columnwidth]{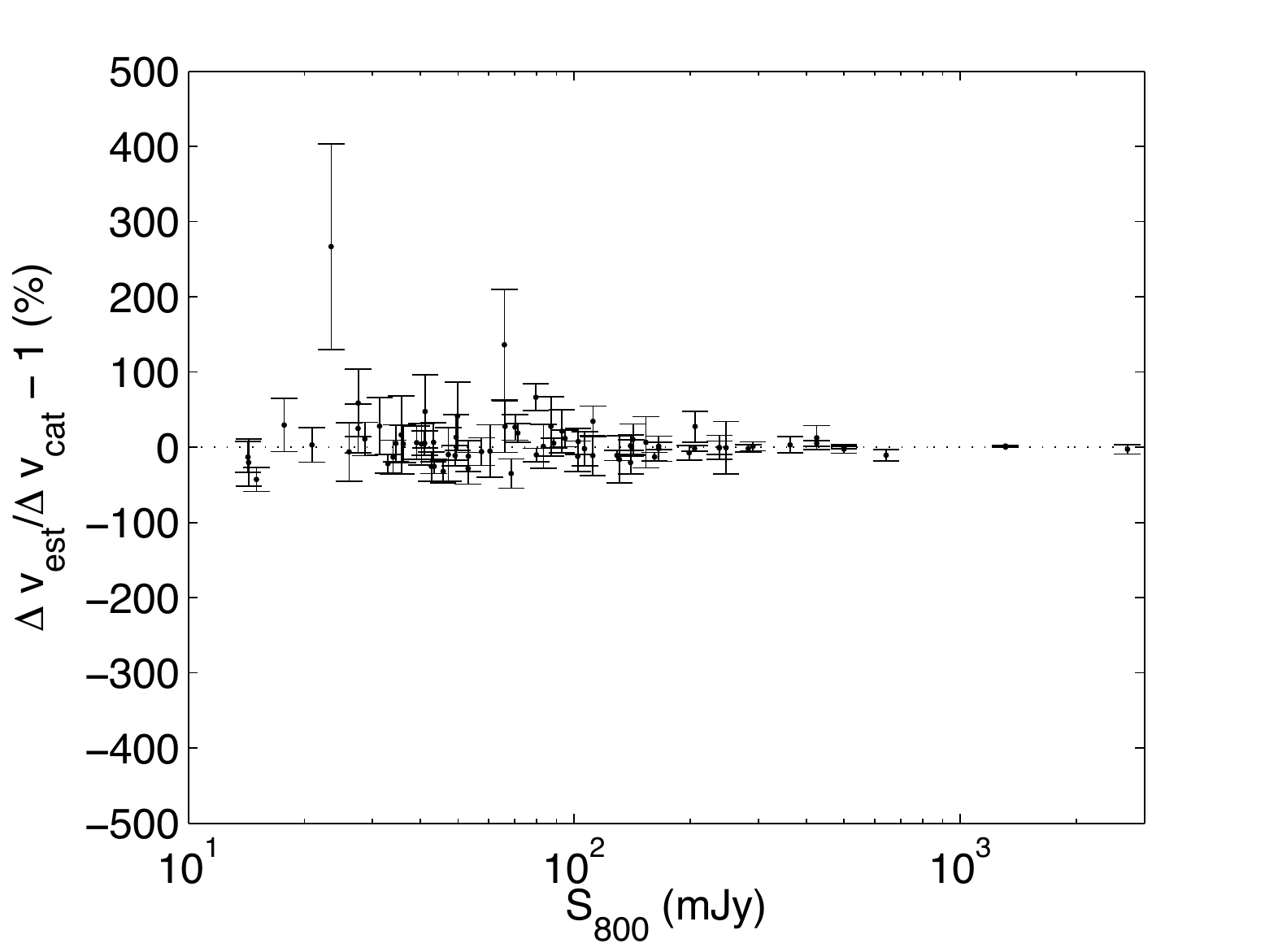}
\caption{Comparison between the estimated and true values for each
  absorption-line parameter, as a function of the 800\,MHz source flux
  density. Sources located outside of the image edge or with $R$ less
  than unity are not included in the sample. The errorbars represent
  the 1\,$\sigma$ uncertainty.}
\label{figure:catalogue_comparison}
\end{figure}

Figure\,\ref{figure:catalogue_comparison} compares the estimated and
input catalogue values for each of the absorption-line model
parameters, plotted as a function of the continuum source flux
density. The redshift position of each line is the most precisely
determined parameter from model fitting, with differences compared to
the input catologue $\sim0.01\,\%$. The peak optical depth and FWHM
parameters are less precisely determined by model fitting to the
simulated data, with differences $\sim10\,\%$. The large majority of
parameters are within $1-3\,\sigma$ of their expected catalogue
values. The few significant outliers are likely due to the imaging
procedure still in development by the ASKAP computing group. We
extract the spectral data from a pixel at the position of the source,
and so either pixelisation of the input model or imaging artefacts may
produce an offset in the estimated parameters.

\subsection{Comparison with \textsc{DUCHAMP}}

We searched for absorption-line components in the same set of 435
simulated ASKAP spectra using the \textsc{Duchamp} 3-dimensional
source finder \citep[for technical details please refer
to][]{Whiting:2008}.

This source finder uses an intensity-thresholding algorithm without
assuming any underlying analytical model for the source shape or line
profile. \textsc{Duchamp} is therefore optimized for detecting complex
sources in 3 dimensions rather than simultaneous detection and
parametric model fitting of spectral-line profiles.

For this work we are considering the spectral domain and so we ran
\textsc{Duchamp} in a mode such that it was optimized for spectral
searches (for example we set the parameters \textsc{searchType} and
\textsc{smoothType} to ``spectral''). The source-finding parameters
are optimised so that we minimise the number of false positive
detections, while maximising the number of detected absorption-line
components. The data are smoothed using a Hanning filter width of 3
and the detections are accepted if they are brighter than a threshold
of $3\,\sigma$ above the mean and contain more than 3 contiguous
channels. In order to correctly determine the total number of
false-positive detections we run the program in both emission and
absorption-line mode.

Of the 540 absorption-line components located within the edge of the
image, 63 are correctly detected with the \textsc{Duchamp} source
finder. We obtain 7 false-positive detections with 3 in absorption and
4 in emission. One of these false positive detections has an
unphysical peak optical depth, while the other 6 have relatively low
SNR and are indistinguishable from the other correctly detected
low-SNR absorption components.

\begin{figure}
\centering
\includegraphics[width = \columnwidth]{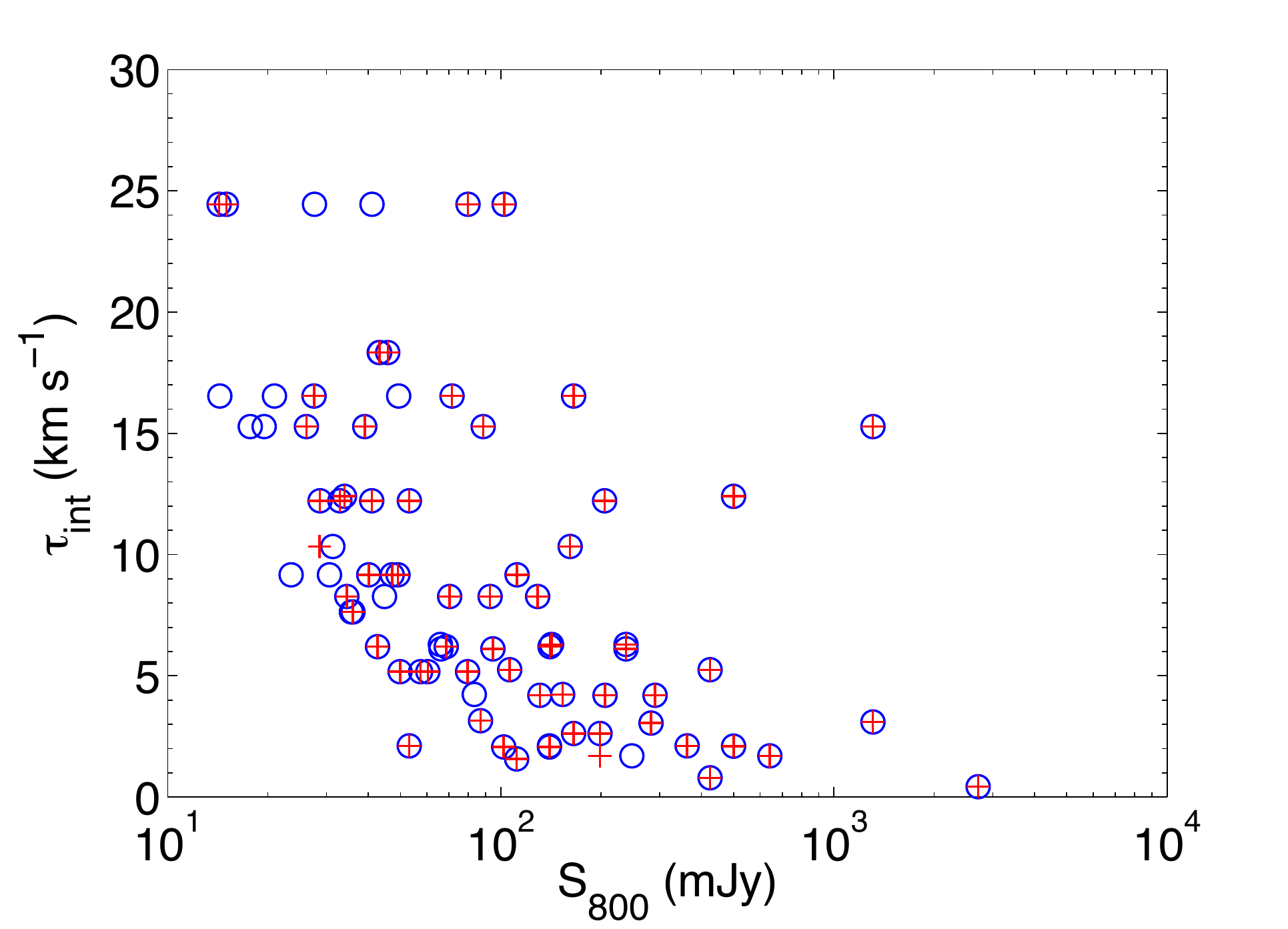}
\caption{The velocity-integrated optical depth versus 800\,MHz source
  flux density for detected absorption-line components using the
  Bayesian line finder (\emph{Blue Circles}) and \textsc{Duchamp}
  (\emph{Red Crosses}).}
\label{figure:finder_comparison}
\end{figure}

Figure\,\ref{figure:finder_comparison} shows the velocity-integrated
optical depth versus the 800\,MHz source flux density for detected
absorption-line components using both the Bayesian line finder and
\textsc{Duchamp}. There are 18 absorption-line components that are
correctly detected with the Bayesian line finder and not with
\textsc{Duchamp}, including 3 which have $R$ less than unity and are
hence rejected due to low significance. Both of the absorption-line
components that are correctly detected with \textsc{Duchamp} and not
with the Bayesian line finder have low SNRs (less than 3) and are
therefore difficult to distinguish from the false positives.

Qualitatively, \textsc{Duchamp} requires significantly lower
computation time for the 1-dimensional spectral-line finding problem,
because calculation of the evidence statistic requires Monte--Carlo
integration over a multi-dimensional parameter space (see
Equation\,\ref{equation:evidence}). However the Bayesian line finder
provides a more robust method for detecting low-significance
spectral-lines, estimating the probability distribution of model
parameters, and selecting between competing analytical models.

\section{Conclusions}
We have applied the multi-nested sampling algorithm to simulated
ASKAP--FLASH data in order to test its usefulness in both finding and
fitting absorption-line components. This Bayesian technique provides
us with a robust tool for selecting spectral-line detections in
low-SNR data, along the line of sight to known continuum sources. The
sampling algorithm is necessarily slower than the \textsc{Duchamp}
source finder because it calculates an estimate of the Bayesian
evidence statistic, and hence provides us with a method of both
assigning significance to our detections and selecting between
competing models. Our analysis of a simulated ASKAP data cube also
shows that the line-finding techniques presented in this paper can
robustly detect HI absorption lines at (and even slightly below) the
levels originally estimated by the FLASH team for a two-hour
integration with ASKAP.

\section*{Acknowledgments} 
We thank the ASKAP computing team for providing the simulated data set
and Farhan Feroz and Mike Hobson for making their \textsc{MultiNest}
software publically available. We also thank B\"{a}rbel Koribalski for
useful discussions and comments. JRA acknowledges support from an ARC
Super Science Fellowship. The Centre for All-sky Astrophysics is an
Australian Research Council Centre of Excellence, funded by grant
CE11E0090.

\bibliographystyle{mn2e.sty}
\bibliography{/Users/jra/reports_thesis_papers/bibliography/trunk/james}

\end{document}